\documentclass[%
preprint,
 amsmath,amssymb,
 aps,
]{revtex4-2}

\usepackage{graphicx}
\usepackage{dcolumn}
\usepackage{bm}
\usepackage{mathtools}
\usepackage{suffix}

\newcommand{\be}{\begin{equation}}
\newcommand{\ee}{\end{equation}}

\UseRawInputEncoding
\DeclarePairedDelimiterX\MeijerM[3]{\lparen}{\rparen}%
{\begin{smallmatrix}#1 \\ #2\end{smallmatrix}\delimsize\vert\,#3}
\newcommand\MeijerG[8][]{%
  G^{\,#2,#3}_{#4,#5}\MeijerM[#1]{#6}{#7}{#8}}
\WithSuffix\newcommand\MeijerG*[7]{%
  G^{\,#1,#2}_{#3,#4}\MeijerM*{#5}{#6}{#7}}

\begin{document}

\title{Fractional discrete vortex solitons}

\author{Cristian Mej\'ia-Cort\'es}
\email{ccmejia@googlemail.com}
\affiliation{Programa de F\'isica, Facultad de Ciencias B\'asicas, 
Universidad del Atl\'antico, Puerto Colombia 081007, Colombia}
\author{Mario I. Molina}
\affiliation{Departamento de F\'isica, Facultad de Ciencias, Universidad 
de Chile, Casilla 653, Santiago, Chile}

\begin{abstract}
    We examine the existence and stability of nonlinear discrete vortex solitons in a
    square lattice when the standard discrete Laplacian is replaced by a fractional
    version.  This creates a new, effective site-energy term, and a coupling among sites,
    whose range depends on the value of the fractional exponent $\alpha$, becoming
    effectively long-range at small $\alpha$ values. At long-distance, it can be shown
    that this coupling decreases faster than exponential: $\sim \exp(- |{\bf
    n}|)/\sqrt{|\bf{n}|}$. In general, we observe that the stability domain of the
    discrete vortex solitons is extended to lower power levels, as the $\alpha$
    coefficient diminishes, independently of their topological charge and/or pattern
    distribution.
\end{abstract}

\maketitle

\section{Introduction}\label{intro} 

Vortices are objects characterized by a spatially-localized distribution of field
intensities, together with a nontrivial phase distribution. This phase circulates around a
singular point, or central core, changing by $2\pi S$ times in each closed loop around it
(where $S$ is an integer number). Integer $S$ is known as the topological charge of the
vortex. The sign of $S$ determines the direction of power flow.  In optics, this type of
solution is also known as a vortex beam and has arisen considerable interest given their
potential technological applications. Optical vortices have been envisioned as a mean to
codify information using their topological charge value in
classical~\cite{PhysRevLett.88.013601} and quantum~\cite{Mair:2001aa} regimes.  Also, a
stable vortex is capable of delivering its orbital angular momentum (OAM) to a nearby
object, given way to one of its most remarkable applications: optical tweezers in
biophotonics, where they are useful due to their ability to influence the motion of living
cells, virus, and molecules \cite{chong_generation_2020,Zhuang188,Favre-Bulle:2019aa}.
Other applications can be found in optical systems communications~\cite{Barreiro:2008aa}
and spintronics~\cite{PhysRevLett.86.4358}. 

A particular domain where discrete vortex solitons can be found, is in the discrete
nonlinear Schr\"{o}dinger (DNLS)
equation~\cite{Kevrekidis2009TheDN,EILBECK1985318,doi:10.1142/9789812704627_0003}, whose
dimensionless form can be written as:
\be
    i {d C_{\bf n}\over{d t}} +  \ \sum_{\bf m} C_{\bf m} + 
    \chi |C_{\bf n}|^2 C_{\bf n} = 0,
\label{eq:1}
\ee
where $C_{\bf n}$ is, for instance, the amplitude of an optical or electronic field,
$\chi$ is the nonlinear coefficient,  and the sum is usually restricted to
nearest-neighbor lattice sites. The DNLS equation has proven useful in describing  a
variety of phenomena in nonlinear physics, such as the transversal propagation of light in
waveguide arrays~\cite{Fleischer:2003aa,LEDERER20081,PhysRevE.70.026602}, propagation of
excitations in a deformable medium~\cite{DAVYDOV1977379,davydov1990}, self-focusing and
collapse of Langmuir waves in plasma
physics~\cite{zakharov1983collapse,zakharov1972collapse}, dynamics of Bose-Einstein
condensates inside coupled magneto-optical
traps~\cite{RevModPhys.78.179,doi:10.1142/S0217984904007190}, and description of rogue
waves in the ocean~\cite{doi:10.1142/S0217984904007190} among others. Its main features
include the existence of localized nonlinear solutions in 1D and 2D,  usually referred to
as discrete solitons, with families of stable and unstable states, the existence of a
selftrapping transition~\cite{MOLINA1993267,TSIRONIS1993135} of an initially localized
excitation, and a degree of excitation mobility in 1D~\cite{PhysRevE.70.026602}.  For the
DNLS equation, the existence and observation of discrete vortex solitons in
Eq.~(\ref{eq:1}) for several  lattices  have been reported in several works. For a square
geometry and Kerr nonlinearity [Eq.~(\ref{eq:1})] it was found that the discrete vortex is
stable when $\chi$ is larger than a critical value~\cite{PhysRevE.64.026601,
PhysRevLett.92.123903, PhysRevLett.102.224102}.  For saturable nonlinearity, discrete
vortices have been experimentally observed in a square
lattice~\cite{PhysRevLett.92.123904}. They have also been studied in a nonlinear
anisotropic Lieb lattice, which possesses a flat band~\cite{Mejia-Cortes:20}. For a
hexagonal lattice in a self-focusing photorefractive crystal, vortices with $S=1$ have
been found but proven unstable, while for $S=2$ a range of stability can be
found~\cite{PhysRevA.79.025801, PhysRevA.83.063825, PhysRevA.79.043821}. Discrete vortices
living at the boundary between a square and hexagonal lattice with photorefractive
nonlinearity,  have also been found~\cite{JOVICSAVIC20151110}.

Another field with substantial recent interest is that of fractional calculus. Its origin
dates back to the firsts observations that the usual integer-order derivative could be
exended to a fractional-order derivative, that is, $(d^n/dx^n)\rightarrow
(d^\alpha/dx^\alpha)$, for real $\alpha$, which is known as the fractional exponent. The
field has a long history  dating back to letters exchanged between L'Hopital and Leibnitz,
followed by later contributions by Euler, Laplace, Riemann, Liouville, and Caputo, to name
some. Several formalisms have been derived to treat these fractional derivatives, each one
having its advantages and shortcomings. In the popular Riemann-Liouville
formalism~\cite{doi:10.1142/8934,west2012physics,miller1993introduction,landkof1972foundations},
the $\alpha$-th derivative of a function $f(x)$ can be formally expressed as
\be
    \left({d^{\alpha}\over{d x^{\alpha}}}\right) f(x) = {1\over{\Gamma(1-\alpha)}} {d\over{d x}}
    \int_{0}^{x} {f(x')\over{(x-x')^{\alpha}}} dx',
    \label{eq:2}
\ee
for $0<\alpha<1$. For the case of the laplacian operator $\Delta=\partial^2/\partial x^2 +
\partial^2/\partial y^2$, its fractional form $(-\Delta)^\alpha$ in two dimensions can be
expressed as~\cite{landkof1972foundations}
\be
(-\Delta)^\alpha f({\bf x}) = 
L_{2,\alpha}  \int { f({\bf x})-f({\bf y})\over{|{\bf x}-{\bf y}|^{2 + 2 \alpha}} }dy,
\label{eq:3}
\ee
with,
\be
L_{2,\alpha} = {16 \ \Gamma(1+\alpha)\over{\pi\  |\Gamma(-\alpha)|}},
\ee
where $\Gamma(x)$ is the Gamma function and $0<\alpha<1$ is the fractional exponent.

The fractional Laplacian (\ref{eq:3}) has found many applications in fields as diverse as
Levy processes in quantum mechanics~\cite{CUFAROPETRONI2009824}, photonics~\cite{Yao:18},
fractional kinetics and anomalous
diffusion~\cite{sokolov2002fractional,ZASLAVSKY2002461,caffarelli2010drift}, strange
kinetics~\cite{Constantin:2016aa},  fluid
mechanics~\cite{caffarelli2010drift,Constantin:2016aa}, fractional quantum
mechanics~\cite{PhysRevE.62.3135,PhysRevE.66.056108}, plasmas~\cite{allen2015boundary},
electrical propagation in cardiac tissue~\cite{doi:10.1098/rsif.2014.0352} and biological
invasions~\cite{Berestycki:2013aa}.

In this work, we study the effect of replacing the usual two-dimensional discrete
Laplacian by its fractional form~\cite{MOLINA2020126180, MOLINA2020126835}, on the
creation and stability of discrete vortex solitons on a square lattice.  As we will see,
as the fractional exponent decreases, moving away from $\alpha=1$, there is a stabilizing
effect on this kind of helical modes, i. e., the power threshold becomes reduced.

\section{Model}\label{model} 

Let us consider a square lattice, where the kinetic energy term in Eq.(\ref{eq:1}),
$\sum_{\bf m}C_{\bf m}$, can be written as $4  C_{\bf n} + \Delta_{n} C_{\bf n}$, where
$\Delta_{n}$ corresponds a the well-known expression for the discretized Laplacian
\begin{equation} 
\Delta_{n} C_{\bf n}= C_{p+1,q}+C_{p-1,q}-4\ C_{p,q} + C_{p,q+1}+C_{p,q-1},
\end{equation}
where ${\bf n}=(p,q)$. Equation (\ref{eq:1}) can then rewritten as 
\be
    i {d C_{\bf n}\over{d t}} + 4 C_{\bf n} + \Delta_{n} C_{\bf n} + \chi |C_{\bf n}|^2 C_{\bf n} = 0.
\label{eq:6}
\ee
Let us now replace the Laplacian $\Delta_{n}$ by its fractional form
$(\Delta_{n})^\alpha$, and given by \cite{CIAURRI2018688,luz2020}
\be 
    (\Delta_{n})^\alpha C_{\bf n} = \sum_{{\bf m}\neq{\bf n} } (C_{\bf m} - 
    C_{\bf n}) \ K^{\alpha}({\bf n}-{\bf m})
\label{eq:7}
\ee
where,
\be 
K^{\alpha}({\bf m}) = {1\over{|\Gamma(-\alpha)|}}\ \int_{0}^{\infty} e^{-4 t }\ I_{m_1}(2 t)\ I_{m_2}(2 t)\ t^{-1-\alpha}\ dt \label{eq:8}
\ee
with ${\bf m} = (m_1,m_2)$ and $I_{m}(x)$ is the modified special Bessel function.
An equivalent expression for $(\Delta_{n})^\alpha$ is
\begin{equation} 
    \begin{split}
    & (\Delta_{n})^\alpha C_{\bf j} =  \\
    & L_{2,\alpha}\sum_{{\bf m}\neq {\bf j}} (C_{\bf m}-C_{\bf j})\
    \MeijerG*{2}{2}{3}{3}{1/2,-(j2-m2+1+\alpha,j2-m2+1+\alpha)}{1/2+\alpha,j1-m1, -(j1-m1)}{1},
    \end{split}
    \label{eq:9}
\end{equation}
where ${\bf j}=(j_1,j_2)$ and ${\bf m}=(m_1,m_2)$, and $G(...)$ is the  Meijer G-function.
As we can see, the symmetric kernel $K^{\alpha}({\bf m})=K^{\alpha}({-\bf m})$ plays the
role of a long-ranged coupling. Near $\alpha=1$, $K({\bf m})\rightarrow \delta_{{\bf
m},{\bf u}}$ where ${\bf u}=(1,0)$ or ${\bf u}=(0,1)$, i,e., coupling to nearest
neighbors only. Equation (\ref{eq:6}) can now be written as
\be
i {d C_{n}\over{d t}} + 4 C_{\bf n} + \sum_{{\bf m}\neq {\bf n}} (C_{\bf m} - C_{\bf n}) K^{\alpha}
({\bf m}-{\bf n}) + \chi |C_{\bf n}|^2 C_{\bf n} = 0,\label{eq:10}
\ee
With a bit of algebraic manipulations, it is possible to prove that Eq.~(\ref{eq:10}) has
two conserved quantities namely,  the power 
\be
P = \sum_{\bf n} |C_{\bf n}(t)|^2
\ee
and the Hamiltonian,
\begin{eqnarray}
    \nonumber & H  = \sum_{\bf n} ( 4-\sum_{{{\bf m}\neq {\bf n}}}K^{\alpha}({\bf n}-{\bf m}))|C_{\bf n}|^2  
    \\
    & \sum_{\bf n}\sum_{{\bf m}\neq {\bf n}} K^{\alpha}({\bf n}-{\bf m}) C_{\bf n}^{*} C_{\bf m} + 
    (\chi/2)\sum_{\bf n} |C_{\bf n}|^4.
\end{eqnarray}
These relations prove useful when monitoring the accuracy of numerical computations. 

\begin{figure}[t!]
    \includegraphics[width=1\columnwidth]{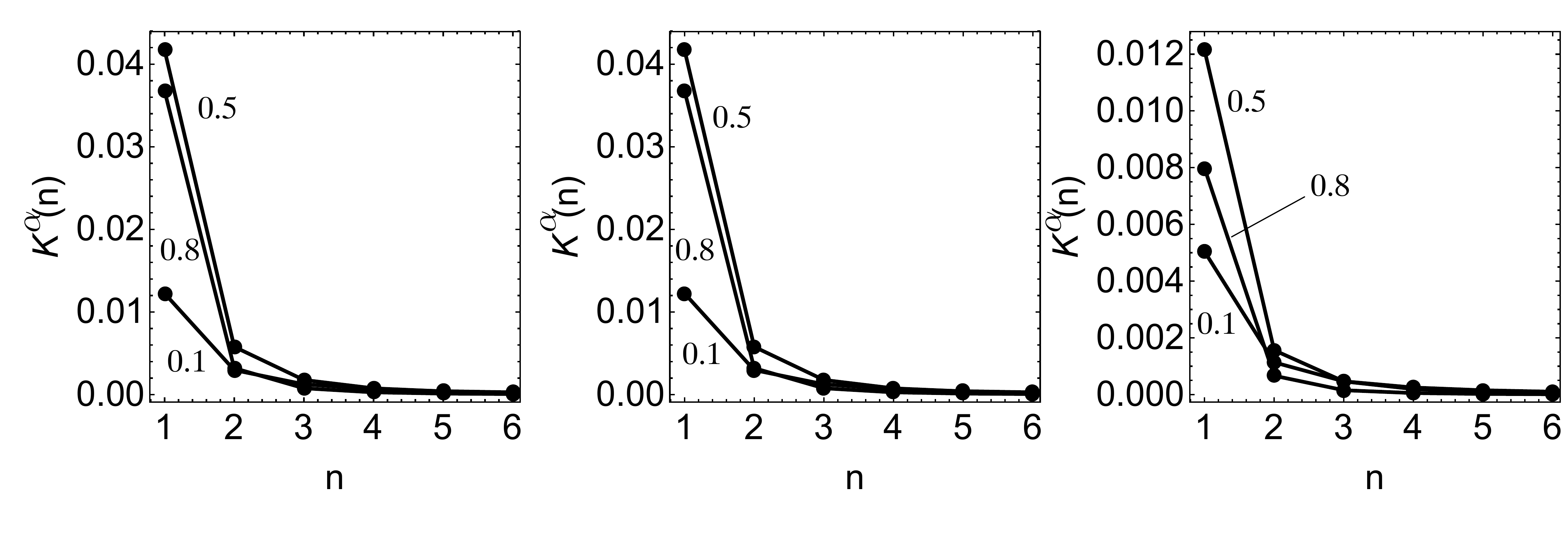}\\
    \includegraphics[width=1\columnwidth]{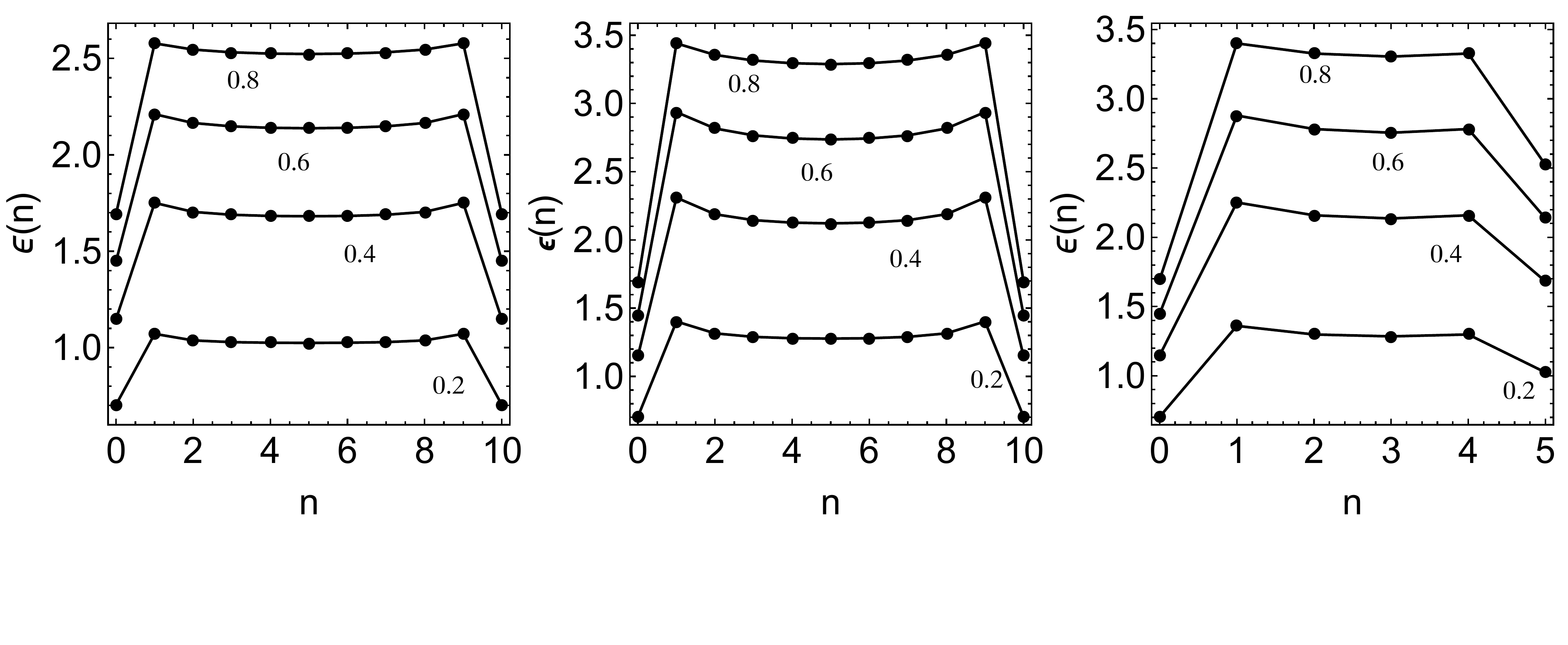}
    \caption{Top row: Effective coupling  $K^{\alpha}({\bf n}-{\bf m})$ between ${\bf
    m}=(0,0)$ and sites ${\bf n}=(n,0)$ (left column),  ${\bf n}=(n,n)$ (middle column),
    and ${\bf n}=(n,2 n)$ (right column). Bottom row: Effective site energy
    $\epsilon({\bf n})$ for several fractional exponents $\alpha$ and ${\bf n}=(n,0)$
    (left column),  ${\bf n}=(n,n)$ (middle column), and ${\bf n}=(n,2 n)$ (right column).
    Number of sites $=10\times 10$.  The numbers on each curve denote the value of the
    fractional exponent.}
    \label{fig0}
\end{figure}

Now let us consider stationary modes defined by $C_{\bf n}(t)=e^{i \lambda t}\ \phi_{\bf n}$, which obey
\be
    (-\lambda + 4 ) \phi_{\bf n} + \sum_{{\bf m}\neq {\bf n}} (\phi_{\bf m} - 
    \phi_{\bf n}) K^{\alpha} ({\bf m}-{\bf n}) + \chi |\phi_{\bf n}|^2 \phi_{\bf n} = 0, 
\label{stationary}
\ee
where $\phi_{\bf n}$ is the field amplitude that defines a complex spatial profile of the
solution, and $\lambda$ is the (eigenvalue) propagation constant. It should be mentioned
that, when dealing with a finite square lattice, in expressions (\ref{eq:6}) and
(\ref{stationary}) the term $4$ is to be replaced by $3\ (2)$ when ${\bf n}$ falls at the
edge (corner).  Figure~\ref{fig0} shows the effective site energy $\epsilon({\bf
n})=4-\sum_{{\bf m}\neq {\bf n}} K^{\alpha}({\bf m}-{\bf n})$ and effective coupling
$K^\alpha({\bf m}-{\bf n})$.  We can see that, as $\alpha$ decreases, the range of the
coupling between two distant sites increases. In particular, for ${\bf n}=0$ and along the
main diagonal ${\bf m}$=$(m,m)$, its value can be shown to approach 
\be K^\alpha(m) \sim
{1\over{|\Gamma(-\alpha)|}} {2^{-2 m}\over{\sqrt{m}}}\hspace{0.5cm} (n\rightarrow
\infty),
\ee
i.e., faster than exponential.

\begin{figure}[t!]
    \includegraphics[width=1\columnwidth]{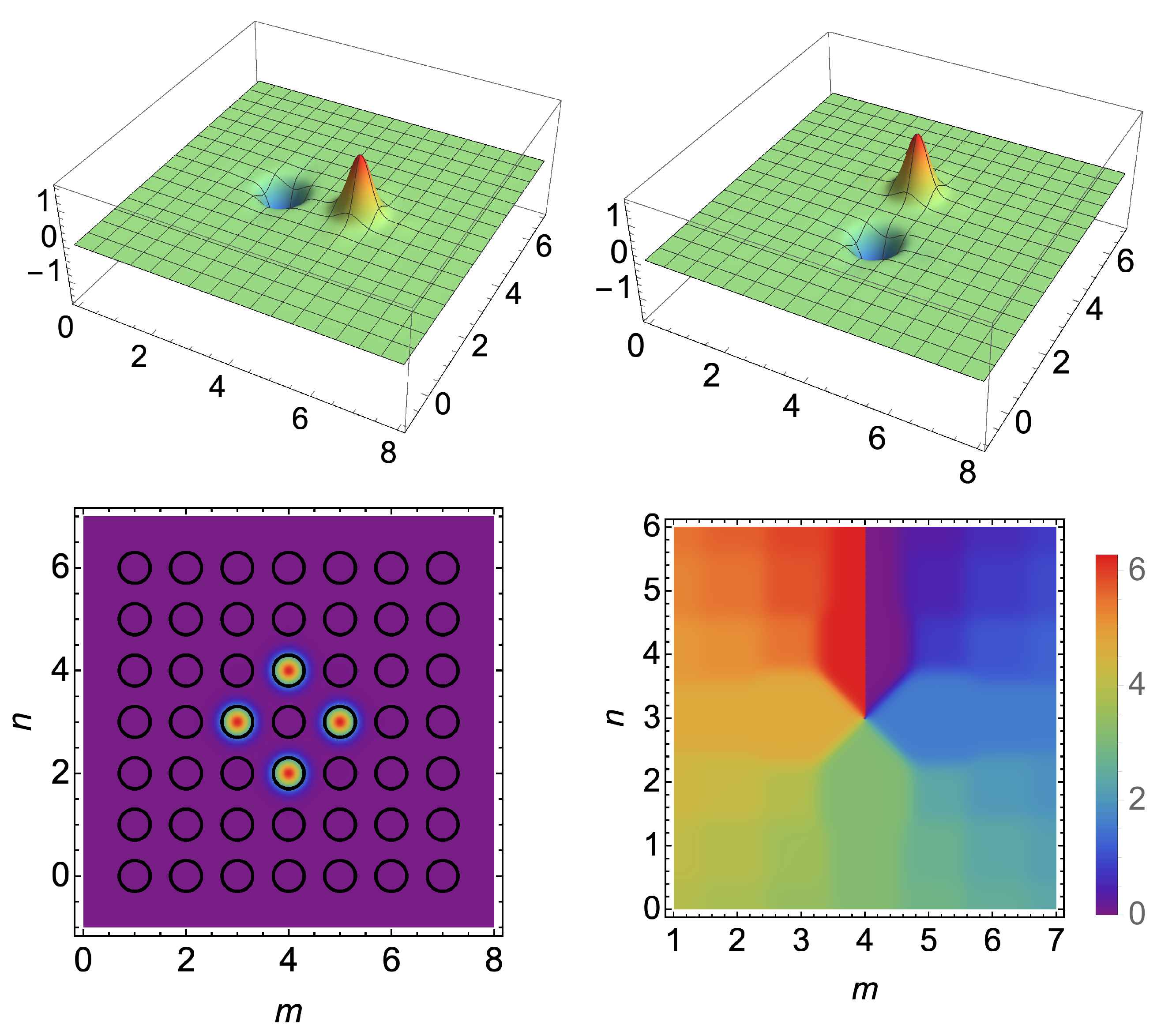}
    \caption{4-sites discrete vortex with $S = 1$ and exponent $\alpha=0.2$.
    Top left: Real part. Top right: Imaginary part. Bottom left:  Amplitude profile. 
    Bottom right: Phase profile. ($\lambda = 6$)}
    \label{fig1}
\end{figure}

\begin{figure}[b!]
    \includegraphics[width=1\columnwidth]{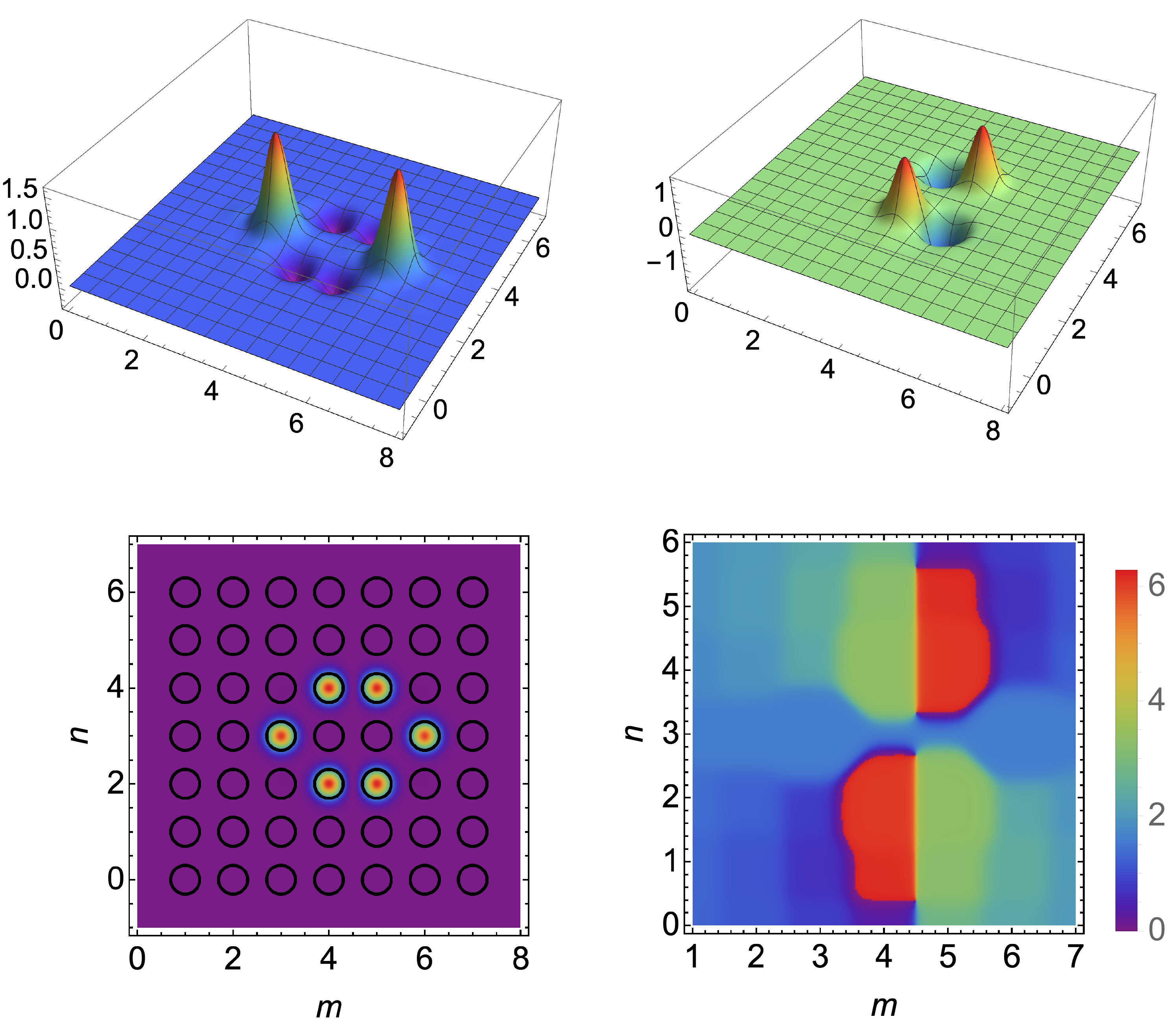}
    \caption{6-sites discrete vortex with $S=2$  and exponent $\alpha=0.2$.  Top left:
    Real part. Top right: Imaginary part. Bottom left: Amplitude profile.  Bottom right:
    Phase profile. ($\lambda=6$) }
    \label{fig2}
\end{figure}

\section{Discrete vortex solitons}\label{results} 

Let us examine the nonlinear stationary modes given as complex solutions of
Eq.~(\ref{stationary}) and characterized by a nontrivial distribution of the phases. They
form a set of $N\times N$ nonlinear algebraic equations for the amplitudes $\{\phi_{\bf n}\}$.
The form of the nonlinear term chosen here is of the Kerr type (cubic), although other
forms can be used, such as the saturable nonlinearity~\cite{Mejia-Cortes:20}.  Numerical
solutions are obtained by the use of a multidimensional Newton-Raphson scheme, using as a
seed a solution in the form $\phi_{n} = A_{n} \exp(i S \theta_{n})$, where $S$ is the
topological charge and $\theta_{n}$ is the azimuthal angle of the $n$th site, with a
highly localized distribution for $A_{n}$. This ansatz is obtained from the decoupled
limit, also known as the anticontinuous limit, where each site becomes decoupled from
each other.  We use a finite $N\times N$ lattice with open boundary conditions.  Figures 2
and 3 show examples of two different discrete vortex solitons with fractional exponent
$\alpha=0.2$, and two values of the topological charge, $S=1$ and $S=2$. The stability of
the computed vortex solitons is carried out by a simple linear stability
analysis~\cite{PhysRevA.86.023834}

\begin{figure}[t!]
    \includegraphics[width=1\columnwidth]{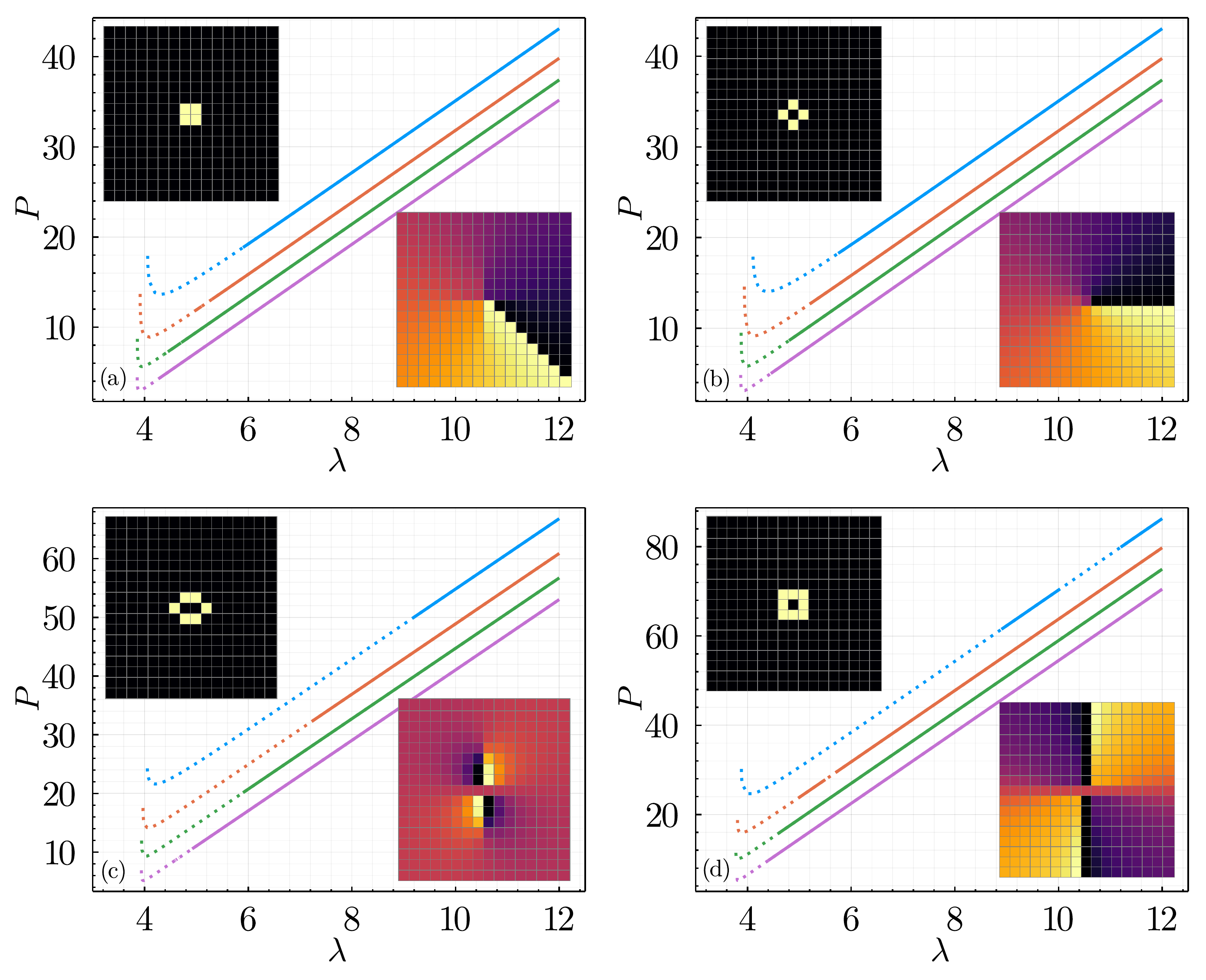}
    \caption{$P$ vs $\lambda$ diagram of some vortex solitons, for
    several fractional exponents and topological charges $S=1$ (upper row) and $S=2$
    (lower row). Solid (dashed) lines represent stable (unstable) solutions. Blue, orange,
    green and violet lines correspond to $\alpha=0.8, \, 0.6, \, 0.4$ and 0.2,
    respectively. Amplitude (top left) and phase profile (bottom right) at inset of
    each diagram corresponds to solutions for $\lambda=12$.}
    \label{fig4}
\end{figure}

Results from the above procedure are displayed in Fig.~\ref{fig4}. They are summarized by
mean of power vs eigenvalue diagrams, for several values of the fractional exponent
$\alpha$.  Amplitude (top left) and phase (bottom right) profiles for these kinds of
stationary vortex solutions are displayed at the inset of each diagram. We see that for
vortex beams with $S=1$ and four main peaks, the off-site square (a) and diamond shape
(b), increase their stability domain as the $\alpha$ coefficient diminish. Similar
behavior can be observed for those stationary modes endowed with $S=2$ and displaying six
main peaks and hexagonal shape (c).  However, for modes with eight peaks and on-site
square shape (d), the stability domain displays a piecewise domain for high values of
$\alpha$. Here, we have employed a $N\times N$ square lattice with $N=17$. As normally
happens in the non-fractional case, families of modes, indistinct of $\alpha$ coefficient,
exhibit a saddle-node bifurcation near to the linear band border. Here we only calculate
solutions belonging to the lower branch of the bifurcation point. Discrete solitons
display here highly localized patterns, as expected for a cubic nonlinearity. We can
observe a smooth spiral phase for any loop enclosing the central core, in those solutions
with symmetric amplitude profiles matching  their nominal topological charge. On the
contrary, for the hexagonal asymmetric pattern, the topological charge only can be
observable in the region where the field amplitude is significant.

In all cases, without exception, the power curves shift down as $\alpha$ is decreased.
Moreover, the main effect of small $\alpha$ values of fractionality is to diminish the power
threshold to obtain stable solutions, which leads to increase the domain of stability of
these helical modes.of these helical modes.  Another observation concerns the limit
$\alpha \rightarrow 0$. In that limit, the range of the coupling diverges and, as a
result, all sites are coupled with each other. Assuming that the amplitude at each site is
nearly identical, the stationary equations (\ref{stationary}) reduce to $(-\lambda + 4 )
\phi + \chi \phi^3 \approx 0$. For $\phi\neq 0$ we have $(4-\lambda)+\chi \phi^2\approx
0$. Using $P\sim Z \phi^2$ where $Z$ is the number of sites initially excited, we have
$P\approx (Z/\chi)(\lambda - 4)$. For $\lambda<4$, we must take $\phi=0$, which implies
$P=0$. This linear dependence can be clearly seen in all plots of Fig.~\ref{fig4} at small
$\alpha$ values.

\section{Conclusions}\label{conclusions} 

In this work we considered the existence and stability of discrete vortex solitons of the
discrete nonlinear Schr\"{o}dinger (DNLS) equation, when the usual Laplacian $\Delta_{\bf
n}$ is replaced by a fractional version $(\Delta_{\bf n})^{\alpha}$ with $0<\alpha<1$. We
employed a square lattice and a Kerr nonlinearity and computed discrete vortex modes and
their stability for different values of the fractional exponent $\alpha$.  Discrete vortex
solitons reported here, namely, the diamond, off and on-site square and hexagonal shape,
exist for any value of fractional exponent and $S=1$ and $S=2$ topological charges.
However, those with diamond and off-site square shape, are only stable for $S=1$. On the
contrary, the on-site square and hexagonal shape cases are stable for $S=2$. The existence
and stability of these modes are strongly related to their spatial distribution,  as well
as to the lattice geometry. In all cases examined, a decrease of the fractional exponent
$\alpha$ causes the power stability curves to shift to lower values, which could be an
intriguing feature since a lower power threshold can ease their experimental observation,
hence, their potential usefulness in photonic applications.

The fractional model that we delve here could be seen as an alternative approach to
describe for example photonic lattices with a long range coupling.  The peculiar effective
long-range coupling could be realized experimentally using a couple waveguide array, by
means of a judicious coupling-engineering~\cite{Vicencio:03}. This kind of optical devices
can be built by femtosecond laser inscription, in amorphous~\cite{Szameit_2010} as well
as crystalline dielectric materials~\cite{Castillo:s}, or in photorefractive crystals,
where the linear refractive index can be modulated externally by light~\cite{Armijo:14}.
In both systems evanescent waves couple to neighbor waveguides determining the transversal
dynamics of light propagation. 

In general, the basic properties of discrete vortices observed  before for the standard
Laplacian exponent ($\alpha=1$) are more or less maintained in the case a fractional
Laplacian. This is itself interesting, since it  suggests that the discrete vortex
soliton properties are robust against mathematical ``perturbations''.

\noindent {\bf Funding}: National Laboratory for High Performance Computing (ECM-02); 
Fondo Nacional de Desarrollo Cient\'ifico y Tecnol\'ogico (1200120).

\noindent {\bf Acknowledgments}: The authors acknowledges helpful
discussions with L. Roncal.

\noindent {\bf Disclosures}: The authors declare no conflicts of interest.





\providecommand{\noopsort}[1]{}\providecommand{\singleletter}[1]{#1}%

\end{document}